# RFID based Automated Car Theft Detection and Arresting system

Anirban Chowdhury, Siddhartha Sarkar, Saikat Das, Dr. Subhasis Bhaumik
School of Mechatronics & Robotics
Bengal Engineering & Science University Shibpur, Howrah, West Bengal, India
Email: 59.anirban@gmail.com[1] ,itssiddharthasarkar@gmail.com[2]
saikat.aeie@gmail.com[3], sbhaumik_besu@yahoo.co.in[4]

*Abstract*—In this paper we are presenting a new approach of car theft detection and arresting system using RFID technology. The main purpose of this paper is to establish the concept and the architecture of the whole system. The system which we are able to develop still now is basically a demonstration model or a prototype of the system. The key thing in this system is the passive RFID tags which will be hidden inside the car and act as a unique identification number for the car. The information of all such tags will be maintained by a centralized server. In case of any car being theft a report should be logged into the server by an authorized user and a network of RFID scanners installed in different check post, traffic signal or toll plazas in the city will search for the reported tag. Once it is found necessary security system would be activated for arresting the car.

## I. INTRODUCTION

The whole concept was developed keeping in mind that the car owner should face very less trouble and expenditure to install a security system which can rescue his car if it is stolen. We know that today RFID technology is used worldwide in many different applications like railway ticketing system, library book management, toll tax collection etc. The main advantage of RFID technology is that a variety of goods or products can be uniquely numbered by RFID tags which are very low cost. And these tags are scanned by a RFID scanner which reads the tag number and sends it to the computer for further processing. So if the car owners purchase RFID tags and keep that in a secret place of the car then his car will have a special number and this number will be registered in a centralized server. This server will keep the status of the tags as they are safe or stolen as reported by the car owner and take necessary action according to that. Thus this system gives the user the freedom and flexibility to the customers to take fewer headaches at any crisis and also to alter his security needs as desired. Customers will also have the facility to change the tags just the change passwords to secure their car more efficiently. One more advantage of this system is that it uses passive tags which require no electricity as it only gets activated when it comes in the radiation range of the scanner.

## II. LITERATURE REVIEW

Mario W. Cardullo introduced RFID technology in 1973 [1]. This is a unique identification system based on electromagnetic coupling in radio frequency range [8]. The technology consists of some tags having small memory where the unique id and some additional information are stored and they are capable to transmitting this information on demand. To interrogate these tags the RFID readers are incorporated [5]. These systems are very rugged in environmental conditions an also useful in tracking objects in motion beyond the line of sight also. The operating frequency of RFID typically lies between 30 KHz to 2.5GHz [6]. The use of RFID tags can be a good alternative in numbering the vehicles [4]. And it can also be used for finding the lost vehicles [7]. Therefore architecture can be thought of where these vehicle tags are being attached to a particular portion of the vehicles in such a way that they can be read through an RFID reader. The information collected by the reader can be managed by several servers connected through LAN or WAN [2]. RFID technology is also being successfully used in traffic signalling [5]. In many places like Philippines RFID has been incorporated in schools and other institutions for access control doors, library management systems and other purposes [3]. Below in fig.1, a typical implementation of RFID system, has been depicted which clearly visualizes the placement of RFID reader and tag in a toll plaza environment for vehicle management system.

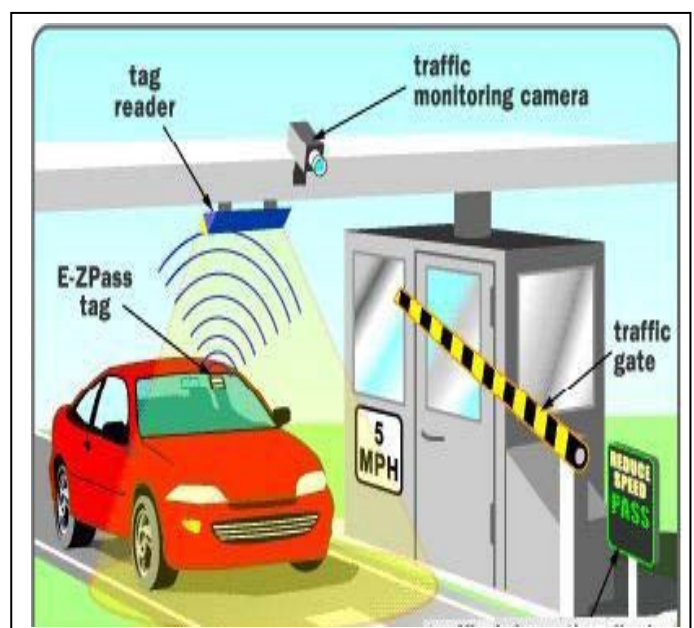

Fig. 1 RFID based vehicle detection system in a toll plaza [ref. 9].

## III. TECHNOLOGY OVERVIEW

Radio Frequency Identification (RFID) is a generic term for non-contacting technologies that use radio waves to automatically identify people or objects. There are several methods of identification, but the most common is to store a unique serial number that identifies a person or object on a microchip that is attached to an antenna. The combined antenna and microchip are called an "RFID transponder" or "RFID tag" and work in combination with an "RFID reader" (sometimes called an "RFID interrogator"). An RFID system consists of a reader and one or more tags. The reader's antenna is used to transmit radio frequency (RF) energy. Depending on the tag type, the energy is "harvested" by the tag's antenna and used to power up the internal circuitry of the tag. The tag will then modulate the electromagnetic waves generated by the reader in order to transmit its data back to the reader. The reader receives the modulated waves and converts them into digital data. There are two major types of tag technologies. "Passive tags" are tags that do not contain their own power source or transmitter. When radio waves from the reader reach the chip's antenna, the energy is converted by the antenna into electricity that can power up the microchip in the tag typically via inductive coupling). The tag is then able to send back any information stored on the tag by modulating the reader's electromagnetic waves. "Active tags" have their own power source and transmitter. The power source, usually a battery, is used to run the microchip's circuitry and to broadcast a signal to a reader. Due to the fact that passive tags do not have their own transmitter and must reflect their signal to the reader, the reading distance is much shorter than with active tags. However, active tags are typically larger, more expensive, and require occasional service. Frequency refers to the size of the radio waves used to communicate between the RFID systems components. Just as you tune your radio to different frequencies in order to hear different radio stations, RFID tags and readers must be tuned to the same frequency in order to communicate effectively. RFID systems typically use one of the following frequency ranges: low frequency (or LF, around 125 kHz), high frequency (or HF, around 13.56 MHz), ultra-high frequency (or UHF, around 868 and 928 MHz), or microwave (around 2.45 and 5.8 GHz). The read range of a tag ultimately depends on many factors: the frequency of RFID system operation, the power of the reader, and interference from other RF devices. Balancing a number of engineering trade-offs (antenna size v. reading distance v. power v. manufacturing cost), the Parallax RFID Card Reader's antenna was designed specifically for use with low-frequency (125 kHz) passive tags with a read distance of around 4 inches. Fig.2 is showing the different components of an RFID reader. Often more than one tag will respond to a tag reader, for example, many individual products with tags may be shipped in a common box or on a common pallet. Collision detection is important to allow reading of data. Two different types of protocols are used to "singulate" a particular tag, allowing its data to be read in the midst of many similar tags. In a slotted Aloha system, the reader broadcasts an initialization command and a parameter that the tags individually use to pseudo-randomly delay their responses. When using an "adaptive binary tree" protocol, the reader sends an initialization symbol and then transmits one bit of ID data at a time; only tags with matching bits respond, and eventually only one tag matches the complete ID string.

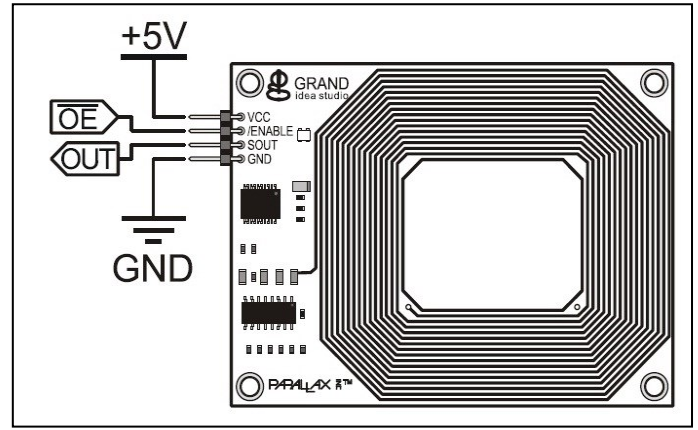

Fig. 2 RFID Reader Module

## IV. COMMUNICATION PROTOCOL

The Parallax RFID Card Reader Serial version easily interfaces to any host microcontroller using only four connections (VCC, /ENABLE, SOUT, GND). Table I describes the use of each of these connections.

TABLE I.

| Pin | Pin Name | Type Function | Function |
|---|---|---|---|
| 1 | VCC | P | System power. +5V DC input. |
| 2 | /ENABLE | I | Module enable pin. Active LOW digital input. Bring this pin LOW to Enable the RFID reader and activate the antenna. |
| 3 | SOUT | O | Serial output to host. TTL-level interface, 2400 bps, 8 data bits, no parity, 1 stop bit. |
| 4 | GND | G | System ground. Connect to power supply's ground (GND) terminal. |

All communication is 8 data bits, no parity, 1 stop bit, and least significant bit first (8N1) at 2400 bps. The RFID Card Reader Serial version transmits data as 5V TTL-level, non-inverted asynchronous serial. The RFID Card Reader USB version transmits the data through the USB Virtual COM Port driver. This allows easy access to the serial data stream from any software application, programming language, or interface that can communicate with a COM port. When the RFID Card Reader is active and a valid RFID transponder tag is placed within range of the activated reader, the tag's unique ID will be transmitted as a 12-byte printable ASCII string serially to the host. The start byte and stop byte are used to easily identify that a correct string has been received from the reader (they correspond to line feed and carriage return characters, respectively). The middle ten bytes are the actual tag's unique ID. For example, for a tag with a valid ID of 0F0184F07A, the following bytes would be sent: 0x0A, 0x30, 0x46, 0x30, 0x31, 0x38, 0x34, 0x46, 0x30, 0x37, 0x41, 0x0D. As the two bits, one is the start bit(=0) and the other is the stop bit(=1) is added with every byte that is transmitted serially, as overhead bits, as per UART protocol, which is marked as blue(start),& red(stop) below. The start byte is always 0A (hex) = 0000010101(binary). The end byte is always   0D (hex) = 00000110111(binary). The 10 digit unique identification number would be like this= (0F0184F07A). The detailed description of the 10 digits is given below:-

- 0(hex)=00011 00001 (ASCII of 0=30(hex))
- F(hex)=00100 01101 (ASCII of F=46(hex))
- 0(hex)=00011 00001 (ASCII of 0=30(hex))
- 1(hex)=00011 00011 (ASCII of 1=31(hex))
- 8(hex)=00011 10001 (ASCII of 8=38(hex))
- 4(hex)=00011 01001 (ASCII of 4=34(hex))
- F(hex)=00100 01101 (ASCII of F=46(hex))
- 0(hex)=00011 00001 (ASCII of 0=30(hex))
- 7(hex)=00011 01111 (ASCII of 7=37(hex))
- A(hex)=00100 00011 (ASCII of A=41(hex))

N: B: The ASCII code corresponding to the digit is transmitted. Now the total serial bit stream would be of 12(number of bytes) x 10(bits in each frame) =120bits. It is shown below:-
00011 00001 00100 01101 00011 00001 00011 00011 00011 10001 00011 01001 00100 01101 00011 00001 00011 01111 00100 00011.

## V. OUR OBJECTIVE

Our main objective was to build a concept and implement in a small scale for visualization. As this kind of systems needs huge resources and cutting edge technologies which are still not available commercially, we decided it to implement it in a lab environment using a wheeled mobile robot and a short range RFID scanner ranging 4 inch. We have made small RFID scanner check post which is able to scan the tag hidden in the mobile robot while it passes through the check post. There is a microcontroller circuit interfaced with the reader which receives the bytes sent by the RFID reader and then transmit them to a standalone PC (Personal Computer) for further processing. Application software is running in the PC which checks the validity of the received tag to identify whether the car is stolen. On encountering an invalid tag it sends a code to the microcontroller circuit which then activates the necessary security system to trap the car inside the check post. Here we represented the car with the mobile robot which follows a white line on a black arena.

## VI. VIEW OF THE ENTIRE ARCHITECTURE

The system we have developed is working in the lab environment. The model we have presented is a reduced model. The system we have developed still now follows the architecture shown in the Fig.3. The only difference is that the entire tag database is stored in the standalone PC but in the real scenario the database would be in a remote server from where the standalone PC at the check post have to access information using TCP/IP protocol.

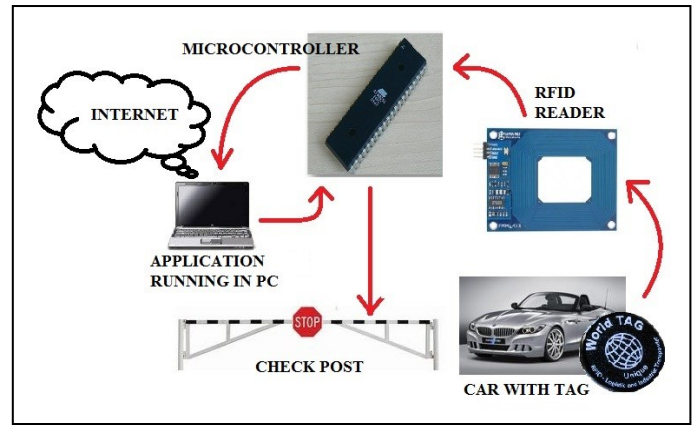

Fig. 3. The System Architecture

## VII. INTRODUCTION OF ASSOCIATED MODULES

We will discuss the components dividing it in some modules.

*A) The car or wheeled mobile robot module:*

The mobile robot is capable of following a white line in a black arena. There are two infrared sensors at the base of the robot chassis to sense the line. The output from the sensor goes to the microcontroller through a comparator circuit which digitizes the analogue data from the sensor. Analysing the output from the sensor the microcontroller generates the motor controlling signals to guide the robot on the right track. There is also a proximity sensor at front portion of the robot which can detect an obstacle within the range of 10 cm. We have put this sensor considering the fact that every car in future would have the obstacle avoidance system inbuilt. This also saves the car from damage when the lock gate of the check post falls suddenly.
Specifications:
Sensors: IR sensors.
Comparator IC: LM339 (four comparisons possible).
Microcontroller: AT89C2051
Motor Driver: ULN2003A (unidirectional motor driver)

*B) The microcontroller module at the check post:*

This module is responsible for receiving the tag's identification number, once it is detected by the scanner. This number is temporarily stored in the microcontroller and sent to the computer. All these communication are serial communication. The communication between RFID reader and the Microcontroller happens at the rate of 2400bps and communication between microcontroller and PC happens in 9600bps. This module also receives the instruction from the computer and controls the gate. The reception from the RFID end and the computer end is controlled by the multiplexer. It switches the connection of microcontroller's Rx pin between PC's Tx pin and RFID reader's SOUT pin.

Specifications:
Microcontroller: AT89S52.
Multiplexer: 74153(1 to 4 multiplexer).
Logic Converter: MAX232 (between RS232 and TTL/CMOS)
Motor Controller: L293D (Bidirectional motor drive).

iii) The PC module: There is a application software running in the standalone PC terminal which connected with several databases for managing the user information and tag information. The main purpose of the program is to check the received tag from the microcontroller end whether it is a valid one or a complaint is been logged by the user. Then it sends the decision code to the microcontroller end for further actions. Also it takes care of adding new users, managing their profile. The API (Application Program Interface) is written in C# and the databases are built in MS-Access.

## VIII. DETAILS OF EACH MODULE

### A) The mobile robot module:

The circuit diagram for controlling the mobile robot is shown in Fig.5. The infrared sensors are calibrated with threshold voltage 0.5V. As the sensors are in black arena the voltage is way beyond 0.5 V and we get logic 0 from the comparator output. And when it is on the white line the voltage generated by the sensor is below 0.5V and the comparator output becomes 1. The threshold for obstacle detection sensor is 4V. Now there is an algorithm running inside the microcontroller which generates the signals to instruct the robot for turning left, right or going forward. The algorithm in the robot car side is very simple. The robot car will go forward by default means when the sensor will sense no obstacle. When the sensor senses an obstacle the microcontroller stops the motor immediately. The flowchart depicting the algorithm is given below in Fig.4. Generally the IR sensors are sensitive to sun light also. So to make it work in day light we need to modulate the transmission frequency of the transmitting diode and shift it away from the frequency of the IR radiation present in the sunlight. A demodulator is also needed at the receiver end to distinguish the frequency and receive it. The other way of sensing these black and white lines is to uses photosensitive registers called LDRs or light dependent registers which generally drops their resistance value in presence of light. But as we have to depend on the reflected light coming from the surface, the sensor modules should be covered properly from stray light sources.

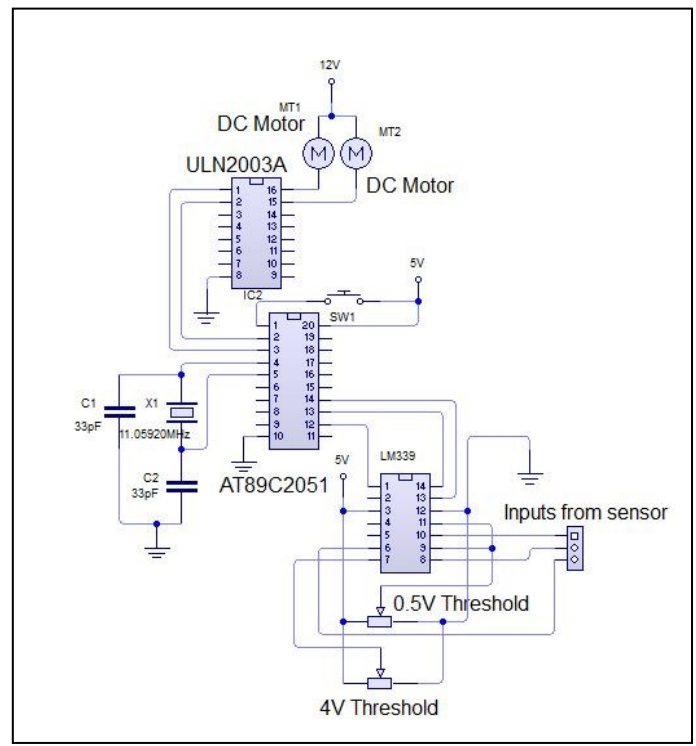

Fig. 5. Circuit Schematic of the mobile robot module

### B) Microcontroller module at the check post:

The microcontroller module at the check post initially receives the 12byte code from the RFID scanner module when the mobile robot passes through the check post. There is a 1 to 4 multiplexer which connects the SOUT line of the RFID reader with the Rx pin of AT89S52. As it receives the bytes, it starts transmitting them to the PC through its Tx pin. Then it changes the select inputs of the multiplexer and connects the Tx line from the PC end with the microcontroller's Rx. It waits for a code to come from the PC end. After receiving the code it interprets it. There are two codes one for arresting the car and one for letting the car to go. The microcontroller

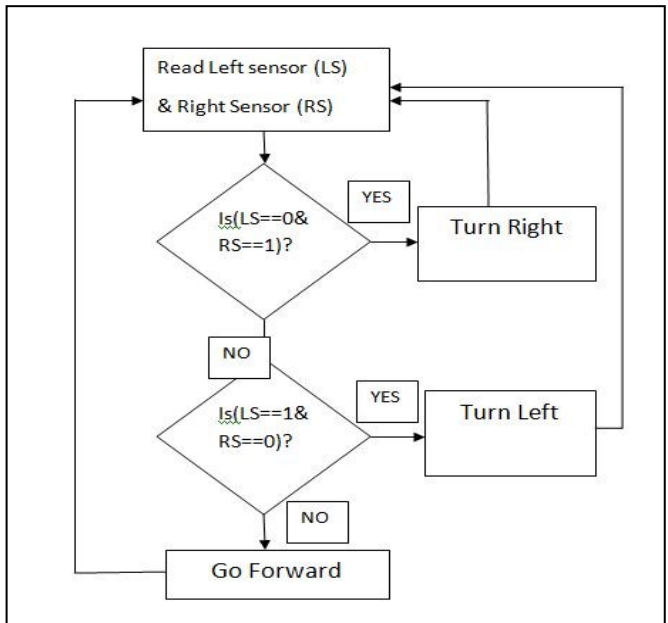

Fig. 4. The line following algorithm

closes the gate as it gets the first code and keeps the gate open as it receives the second code. After receiving the code it sends the same back to the computer to show whether there is any mistake in computation or transmission. Fig.7 is showing the circuit schematic of the module. The algorithm for controlling the whole process is given in steps in the following section. The purpose of the algorithm is to read the RFID tag in continuous mode and as he gets a tag it sends it to the computer. Then it waits for receiving the signal from the computer. After receiving it re-transmits it to the same data to the computer to verify whether the application program running in the computer is working right. Finally it performs the motor action for closing the gate or keeping it open. The algorithm is depicted in Fig.6

*C) Overview of RCTDAS-GUI:*

The RCTDAS system is developed for the beneficial of the car owners of the concerned area which is under the surveillance of RCTDAS. So the whole system should be centralised so that the owners don't have to bother after they have registered once during their purchase or at any later time under the system, moreover system operators at different check-posts should also have the capability to respond to modify the user requests and user registration details with ease. To provide all of the above mentioned features all we needed is a common Graphical User Interface which may work stand alone or under the network of RCTDAS. The authenticity of an user is given much priority here and the system takes care of the user flexibility. The real image of the wheeled mobile robot, the microcontroller circuit at the check post and the view of the entire system we have developed is shown in Fig.8

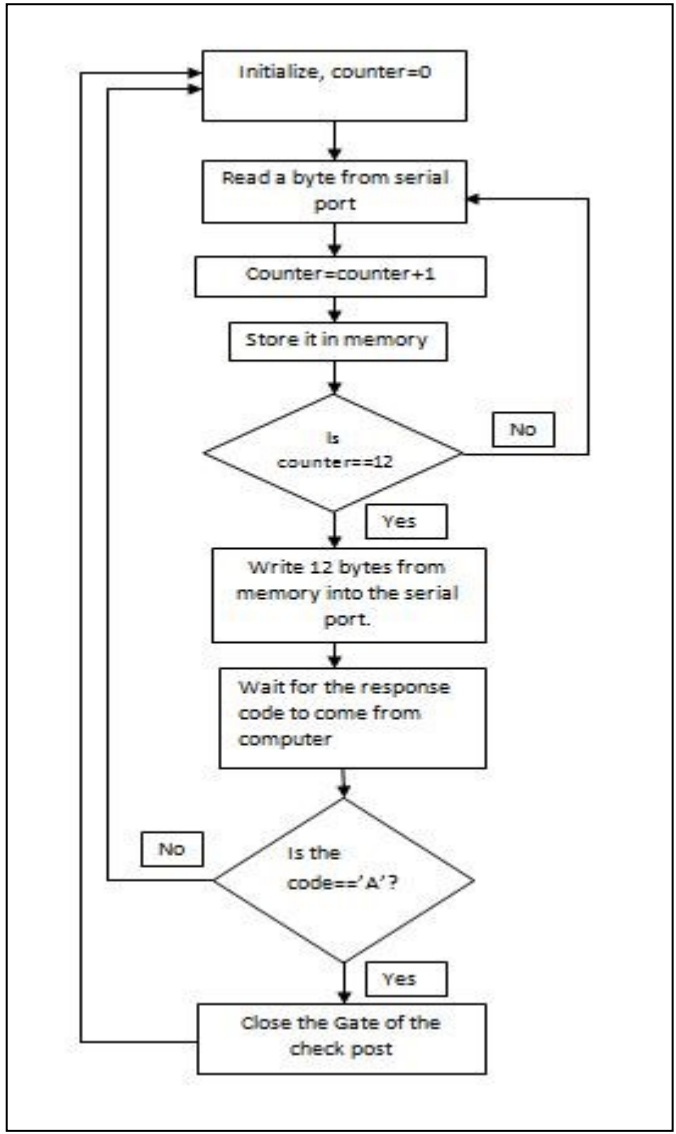

Fig.6 Algorithm running at the microcontroller end

The circuit diagram at the check post side is depicted in Fig.6

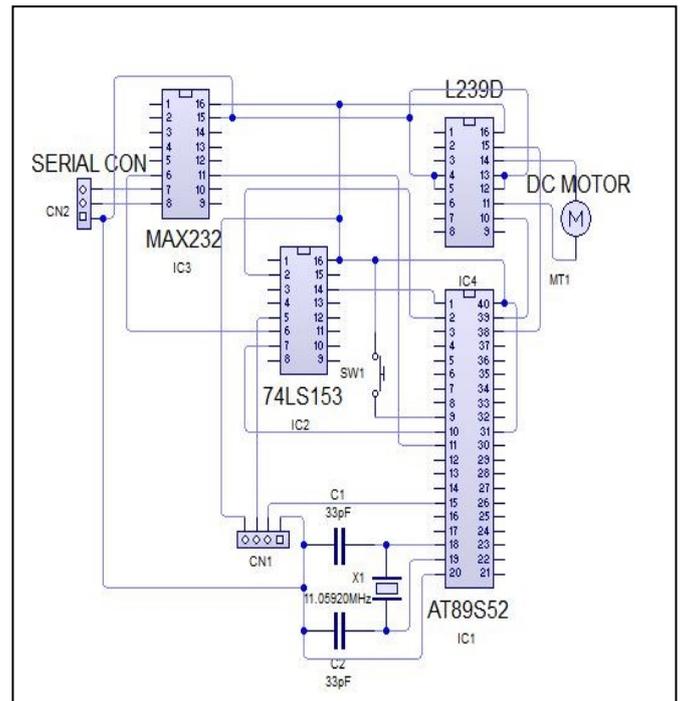

Fig.7 Circuit Schematic for interaction between RFID scanner and PC.

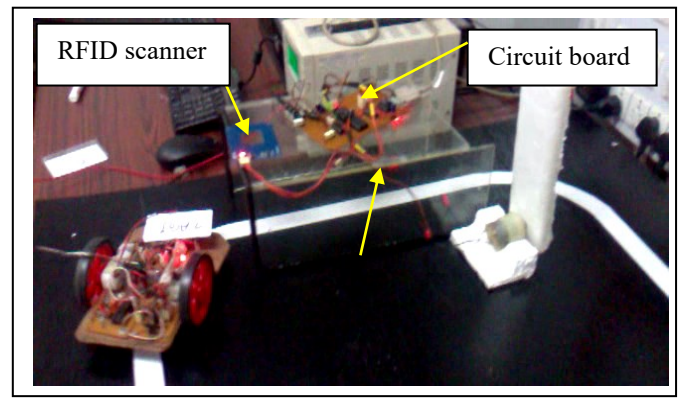

Fig.8 Implementation of the whole system in Lab

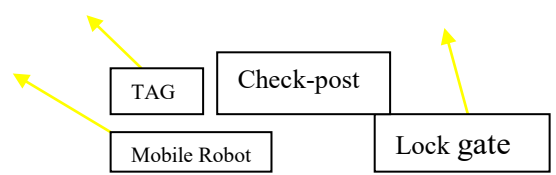

Features of RCTDAS-GUI:
- Light-weight GUI
- No installation required for the Users
- Separate login IDs for Users and Operators
- Different GUI environments for Users and Operators according to their needs
- All the registered car owner will get separate Login Id and Password against each car for increased security
- User can change their account details and password at anytime from anywhere through internet and RCTDAS GUI
- User can report for Missing Car anytime From anywhere through internet and RCTDAS GUI

D) *Development of the API:*

RCTDAS Process-Flow is depicted below in Fig. 9. It is developed using Visual Studio 2010 which is an Integrated Development Environment for C#.net is used as Front End Language and in back end Database is managed through SQL Queries by using OLEDBConnection, OLEDBDataadapter, and OLEDBCommandbuilder. A Split Database is created using MS-Access 2010 and disconnected data model is used so that the same Database can be edited by several owners and operators simultaneously.

Only the front-end part of the split database is distributed along with every copy of RCTDAS-GUI but the backend part is stored in any shared location of the Local Area Network to ensure real-time updating.

IX. CONCLUSION

In this paper all we have done is to build the concept how the system architecture will look like. But this prototype model is implemented in a small lab environment using short range RFID scanners. Here the real challenges are still left like finding a suitable RFID scanner which would cover the range as big as a normal highway so that this system could be implemented real ground. Also these scanners should be installed in different parts of the city outposts creating a network which can be communicated through internet. Thus the initial installation of the whole system would be an issue. Besides it there are some security related issue which needs to be solved before final installation.

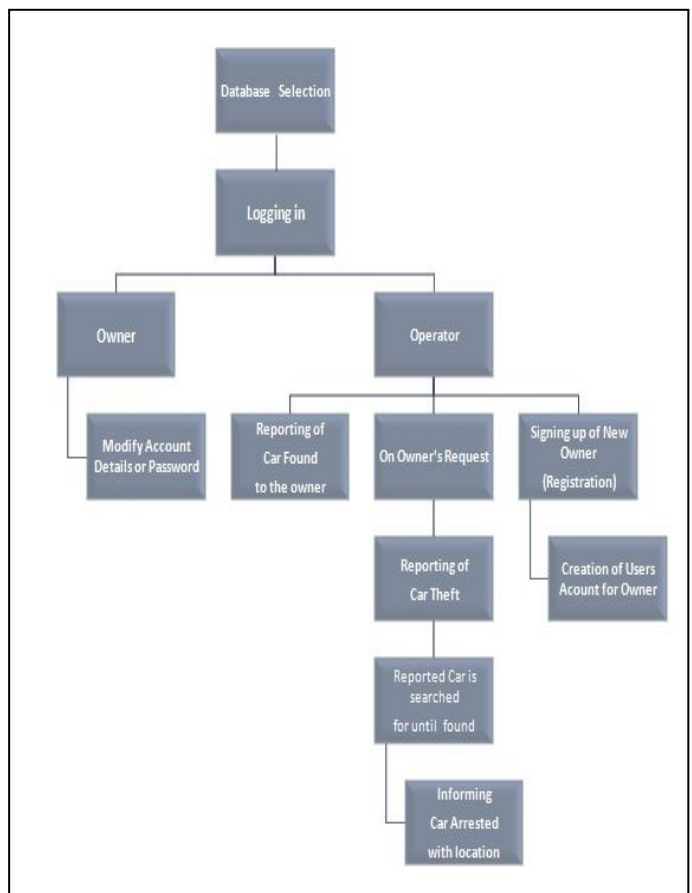

Fig. 9. Architecture of the RACTDAS GUI.